\documentclass{iaus}

\usepackage{graphicx}
\newcommand{\EQ}{\begin{equation}}
\newcommand{\EN}{\end{equation}}
\newcommand{\EQA}{\begin{eqnarray}}
\newcommand{\ENA}{\end{eqnarray}}

\newcommand{\Eq}[1]{equation~(\ref{#1})}

\newcommand{\Fig}[1]{Fig.~\ref{#1}}

\newcommand{\bra}[1]{\langle #1\rangle}

\newcommand{\mean}[1]{\overline #1}

{}
\newcommand{\meanFFFF}{\overline{\mbox{\boldmath ${\cal F}$}}{}}{}

\newcommand{\meanSSSS}{\overline{\mbox{\boldmath ${\mathsf S}$}} {}}

{}
{}
{}
{}
{}
{}
{}
{}
{}
\newcommand{\meanBB}{\overline{\mbox{\boldmath $B$}}{}}{}
{}
{}
{}
{}
\newcommand{\meanUU}{\overline{\mbox{\boldmath $U$}}{}}{}
{}
{}

\newcommand{\meanB}{\overline{B}}

\newcommand{\meanFFF}{\overline{\cal F}}
{}

{}

{}
{}

\newcommand{\yyy}{\hat{\mbox{\boldmath $y$}} {}}

\newcommand{\uu}{\mbox{\boldmath $u$} {}}

\newcommand{\bb}{\mbox{\boldmath $b$} {}}
\newcommand{\BB}{\mbox{\boldmath $B$} {}}

\newcommand{\jj}{\mbox{\boldmath $j$} {}}

\newcommand{\ee}{\mbox{\boldmath $e$} {}}

\newcommand{\nab}{\mbox{\boldmath $\nabla$} {}}

\newcommand{\dd}{{\rm d} {}}

\def\Co{\mbox{\rm Co}}
\def\Sh{\mbox{\rm Sh}}
\def\St{\mbox{\rm St}}

\def\Rm{R_{\rm m}}

\def\Co{\mbox{\rm Co}}

\def\kf{k_{\rm f}}
\def\urms{u_{\rm rms}}

\def\Beq{B_{\rm eq}}

\def\half{{\textstyle{1\over2}}}

\newcommand{\s}{\,{\rm s}}
\newcommand{\mpers}{\,{\rm m/s}}

\newcommand{\Mm}{\,{\rm Mm}}

\newcommand{\yapj}[3]{ #1, {ApJ,} {#2}, #3}

\newcommand{\yapjl}[3]{ #1, {ApJ,} {#2}, #3}
\newcommand{\yapjs}[3]{ #1, {ApJS,} {#2}, #3}
\newcommand{\yan}[3]{ #1, {Astron.\ Nachr.,} {#2}, #3}

\newcommand{\yana}[3]{ #1, {A\&A,} {#2}, #3}

\newcommand{\ygafd}[3]{ #1, {Geophys.\ Astrophys.\ Fluid Dyn.,} {#2}, #3}

\newcommand{\yaraa}[3]{ #1, {ARA\&A,} {#2}, #3}

\newcommand{\yprl}[3]{ #1, {Phys.\ Rev.\ Lett.,} {#2}, #3}

\newcommand{\ynat}[3]{ #1, {Nature,} {#2}, #3}

\newcommand{\ysph}[3]{ #1, {Solar Phys.,} {#2}, #3}

\newcommand{\ypre}[3]{ #1, {Phys.\ Rev.\ E,} {#2}, #3}
\newcommand{\ypnas}[3]{ #1, {Proc.\ Nat.\ Acad.\ Sci.,} {#2}, #3}

\newcommand{\yjour}[4]{ #1, {#2}, {#3}, #4}

\title[Paradigm shifts]{Paradigm shifts in solar dynamo modeling}

\author[Axel Brandenburg]{Axel Brandenburg}

\affiliation{Nordita, Roslagstullsbacken 23, 10691 Stockholm, Sweden}

\pubyear{2008}
\volume{259}  
\pagerange{159--166}
\date{?? and in revised form ??}
\jname{Cosmic Magnetic Fields:\\
From Planets, to Stars and Galaxies}
\editors{K.\ G.\ Strassmeier, A.\ G. Kosovichev \& J.\ E. Beckman, eds.}
\doi{10.1017/S1743921309030403}
\begin{document}

\maketitle

\begin{abstract}
Selected topics in solar dynamo theory are being highlighted.
The possible relevance of the near-surface shear layer is discussed.
The role of turbulent downward pumping is mentioned in connection with
earlier concerns that a dynamo-generated magnetic field would be rapidly lost
from the convection zone by magnetic buoyancy.
It is argued that shear-mediated small-scale magnetic helicity fluxes
are responsible for the success of some of the recent large-scale
dynamo simulations.
These fluxes help in disposing of excess small-scale magnetic helicity.
This small-scale magnetic helicity, in turn, is generated in response to
the production of an overall tilt in each Parker loop.
Some preliminary calculations of this helicity flux are
presented for a system with uniform shear.
In the Sun the effects of magnetic helicity fluxes may be
seen in coronal mass ejections shedding large amounts of
magnetic helicity.
\end{abstract}

\firstsection 
\section{Introduction}

Unlike the geodynamo, which is widely regarded a solved
problem (e.g.\ Glatzmaier \& Roberts 1995), the solar dynamo problem
remains unsolved in that there is no single model that is free of
theoretical shortcomings and that actually reproduces the Sun.
The flux-transport dynamo models (e.g.\ Dikpati \& Charbonneau 1999;
K\"uker et al.\ 2001; Chatterjee et al.\ 2004) are currently the only
models that display some degree of realism, but there are aspects in
their design that are arguably not sufficiently plausible.
One problem is the assumption of a rather low turbulent magnetic
diffusivity that needs to be assumed in a {\it ad hoc} fashion.
On theoretical grounds the turbulent magnetic diffusivity should be
comparable with the turbulent kinematic viscosity (Yousef et al.\ 2003),
which is not the case on these flux-transport dynamo models.

The purpose of this review is to collect recent findings that are
relevant in solving the solar dynamo problem.
On the one hand, there are direct simulations of convective dynamo
action in spherical shells by Brun \& Toomre (2002),
Brun et al.\ (2004, 2006), Browning et al.\ (2006), and Brown et al.\ (2007).
These models are already quite realistic and should ultimately be able
to reproduce the solar dynamo.
Their main shortcomings range from still insufficient resolution (even
though it is already exhausting current capabilities) to the negligence
of certain physical features of the model.
Examples of shortcomings include the still insufficient degree of
stratification as well as the simplified treatment or even the neglect
of surface and tachocline boundary layers.
On the other extreme, there are phenomenologically motivated mean-field
models that are constructed based on their success in reproducing the Sun.
These models tend to utilize a subset of known turbulent transport effects
with {\it ad hoc} amplitudes and prescriptions  of their radial profiles.
Such models are made nonlinear by simple $\alpha$ quenching terms.
However, this type of prescription that is still commonly used in mean
field models does not correctly describe the saturation behavior of
large-scale magnetic fields as known from three-dimensional turbulence
simulations.
This is true even for much simpler models, for example those where the
turbulence is driven at a specific length scale via a forcing function
(Brandenburg 2001).

\section{Negative radial shear?}

\begin{figure}[t!]\begin{center}
\includegraphics[width=.32\textwidth]{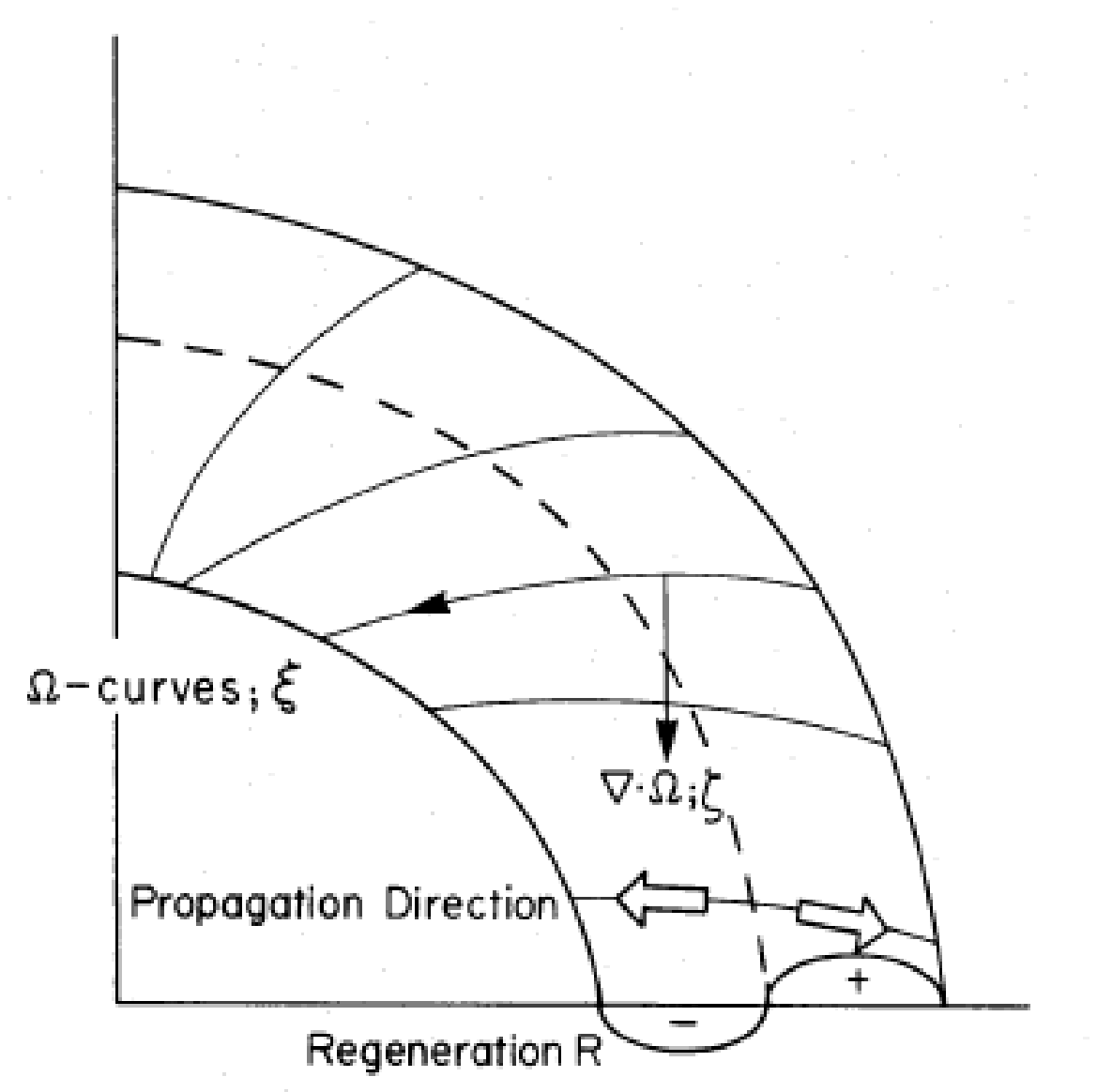}
\includegraphics[width=.32\textwidth]{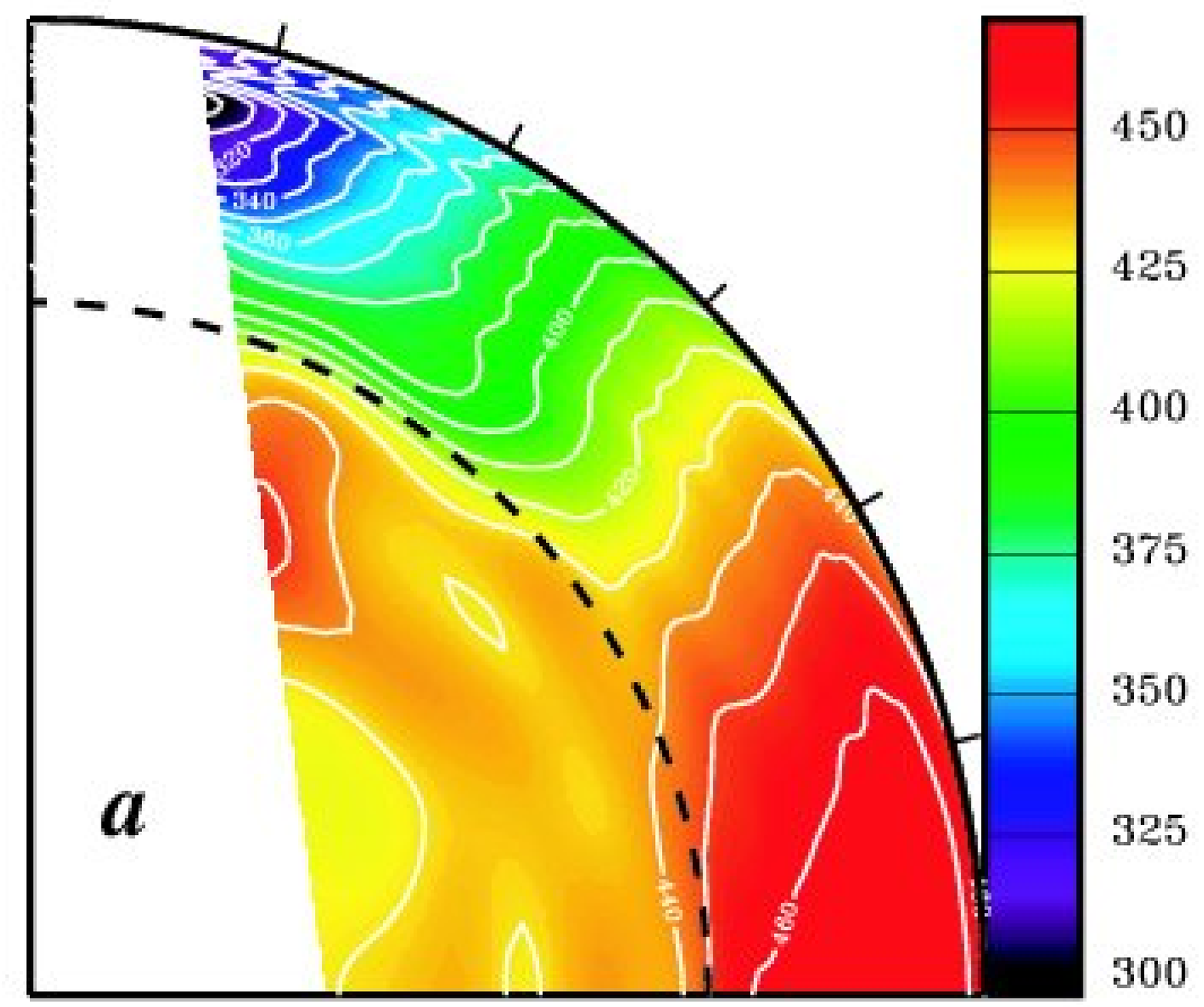}
\includegraphics[width=.34\textwidth]{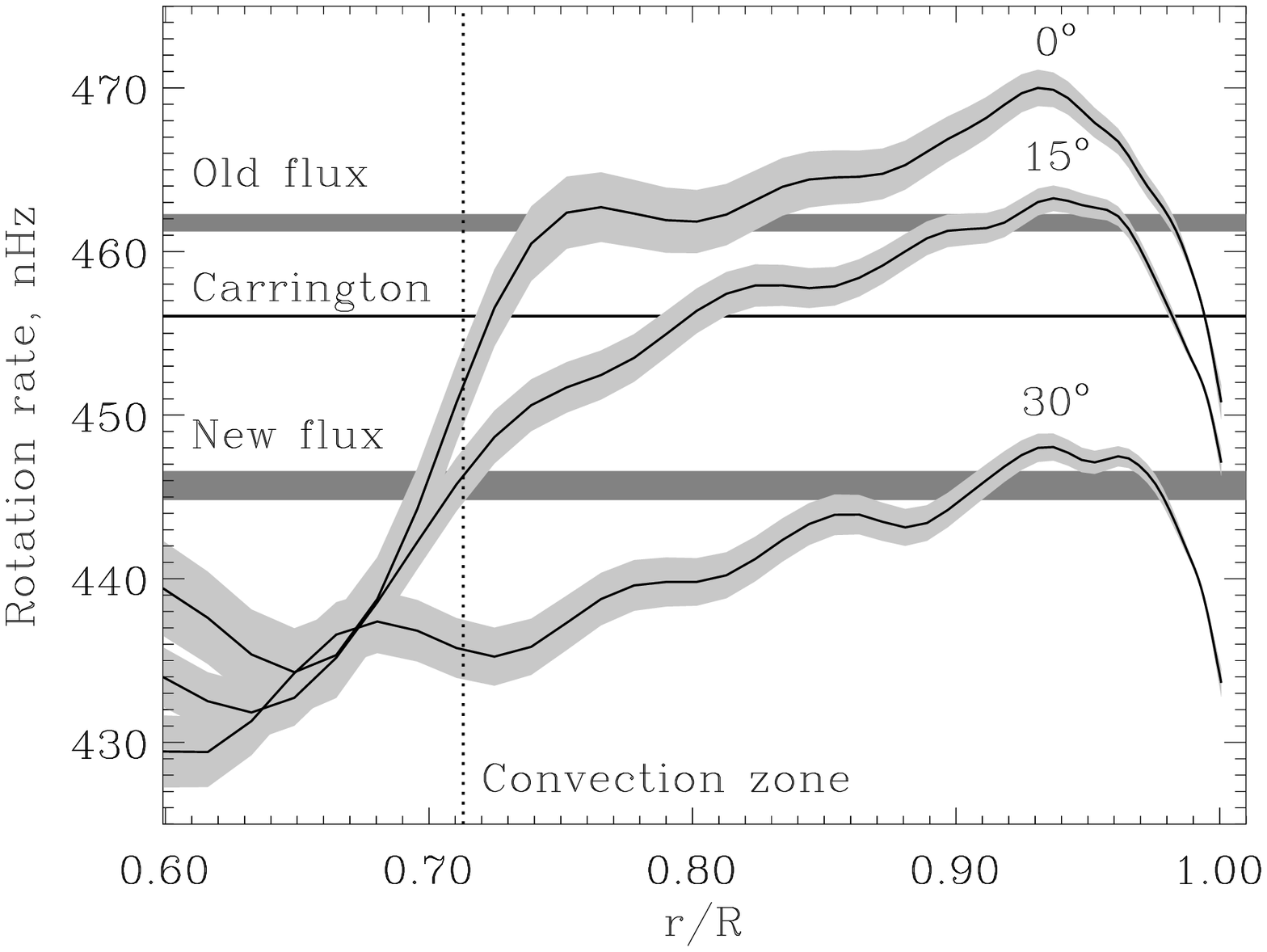}
\end{center}\caption[]{
Comparison of the differential rotation contours that were originally
expected by Yoshimura (1975) based on solar dynamo model considerations
(left) with those by Thompson et al.\ (2003) using helioseismology (middle).
Note the similarities between the contours on the left (over the bulk
of the convection zone) and those on the right (over the outer 5\% of
the solar radius).
In the right hand panel we show radial profiles of angular velocity
as given by Benevolenskaya et al.\ (1999).
Note the sharp negative radial gradient near the surface.
}\label{pdiffrot}\end{figure}

In the 1970s a number of solar dynamo models were discussed that were
based on an $\alpha$ effect that is positive in the northern hemisphere
(and negative in the southern hemisphere)
and a differential rotation that had a negative radial gradient, i.e.\
$\partial\Omega/\partial r<0$ (see the left panel of \Fig{pdiffrot}).
Such models produce an equatorward migration of magnetic activity
toward the equator.
Examples of such models in full spherical geometry include the papers
Steenbeck \& Krause (1969), Roberts \& Stix (1972), K\"ohler (1973),
and Yoshimura (1975).
During the 1980s these models faced two difficulties,
a conceptual one and one based on a conflict with observations.
The conceptual difficulty has to do with the idea that the magnetic
field may be fibral (Parker 1984), i.e.\ it is distributed in the form of many
flux tubes with a small filling factor and relatively large
field strength.
However, such fields would easily be unstable to magnetic buoyancy
instabilities.
This and other considerations led Spiegel \& Weiss (1980) to propose
that the solar dynamo might instead work at the bottom of the solar
convection zone and not, as assumed until them, in a distributed
fashion in the bulk of the convection zone.
In fact, in an earlier paper by Parker (1975) this possibility
was already discussed in some detail.
The conflict with observations had to do with the fact that
helioseismology began to put tight constraints on the form
of the Sun's internal angular velocity profile.
This seemed to rule out a negative radial $\Omega$ gradient.

These two difficulties led to a new type of dynamo model that works
in the lower overshoot layer, where $\alpha$ has the opposite sign
(DeLuca \& Gilman 1986, 1988).
Not only would magnetic flux tubes presumably be stabilized against magnetic
buoyancy instabilities, but at that depth the sign of the $\alpha$ effect
would be reversed and thus dynamo waves would again propagate equatorward
even though $\partial\Omega/\partial r>0$.
However, several problems with this approach have been discussed
(phase relation, number of toroidal flux belts, etc.).
Yet another model is the flux transport model, where a meridional
circulation (equatorward at the bottom of the convection zone) is chiefly
responsible for driving the dynamo wave equatorward.
However, it is now quite clear that there is a strong shear layer
near the surface where $\partial\Omega/\partial r<0$, just as originally
anticipated.
The difference is that now this shear layer only extends over the outer
5\% of the solar radius.
In the past the relevance of this near-surface shear layer was discarded
mainly on the grounds that magnetic buoyancy effects would lead to a
rapid loss of magnetic field.
Another reason is that near the surface the local turnover time is still
short compared with the rotation period.
However, simulations of compressible convection did show some time ago
that magnetic buoyancy is effectively overpowered by the effects of
turbulent downward pumping.
Regarding the timescales, rotational effects are likely to play a role
at the bottom of the near-surface shear layer, which is at a depth of
about $40\Mm$.
The relative importance of rotation and convection is determined by
the Coriolis number,
\EQ
\Co=2\Omega\tau,
\EN
where $\Omega$ is the solar rotation rate and $\tau=H_p/\urms$ is the
turnover time with $H_p$ being the pressure scale height.
Using $\Omega=3\times10^{-6}\s^{-1}$ for the solar rotation rate,
$\urms=50\mpers$ for the rms velocity at a depth of $40\Mm$, and
$H_p=13\Mm$ for the pressure scale height at that depth we find
$\tau=1.3\,\dd$, and hence $\Co=1.4$, suggesting that rotation should
become important at the bottom of the near-surface shear layer.

\begin{figure}[t!]\begin{center}
\includegraphics[width=.49\textwidth]{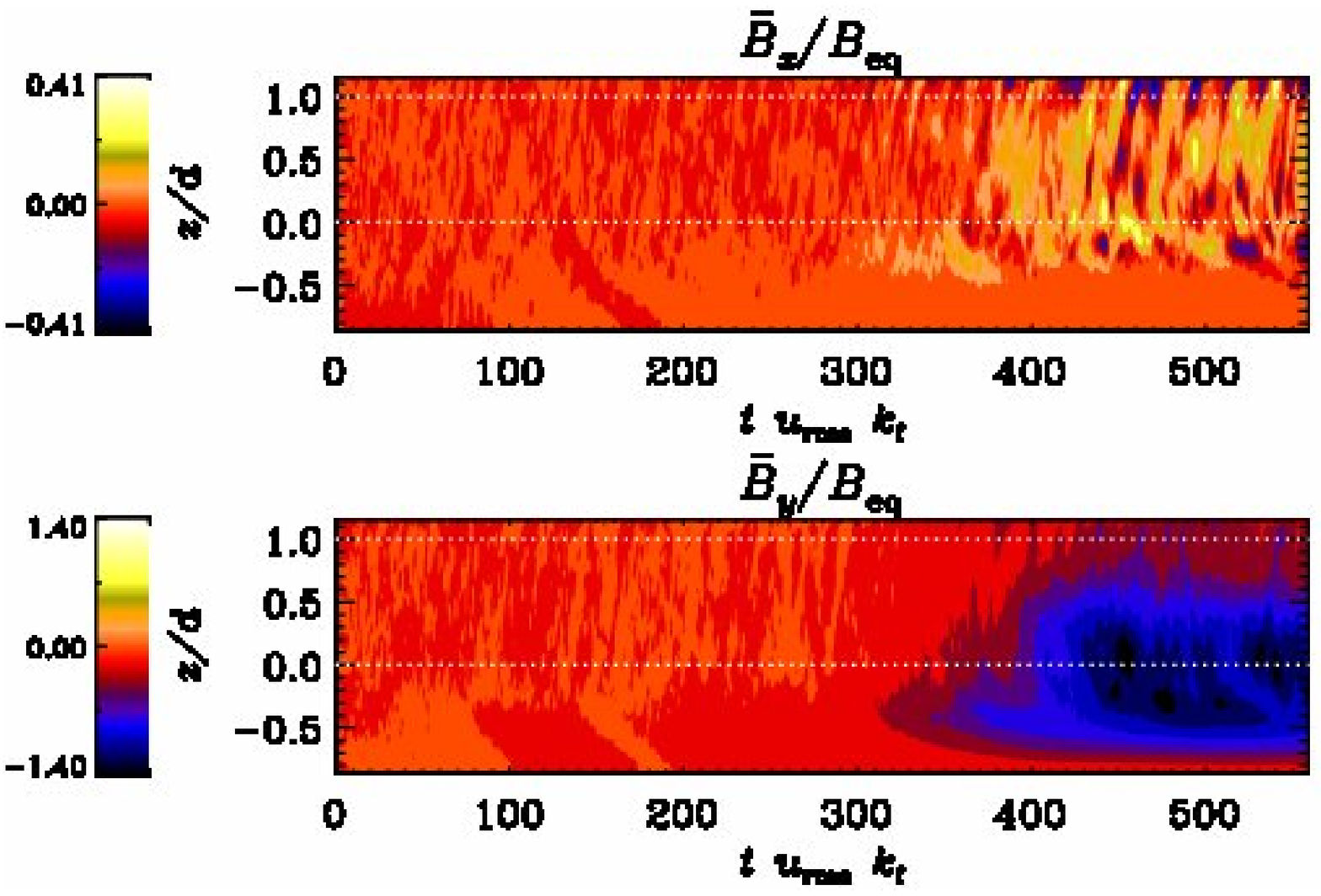}
\includegraphics[width=.50\textwidth]{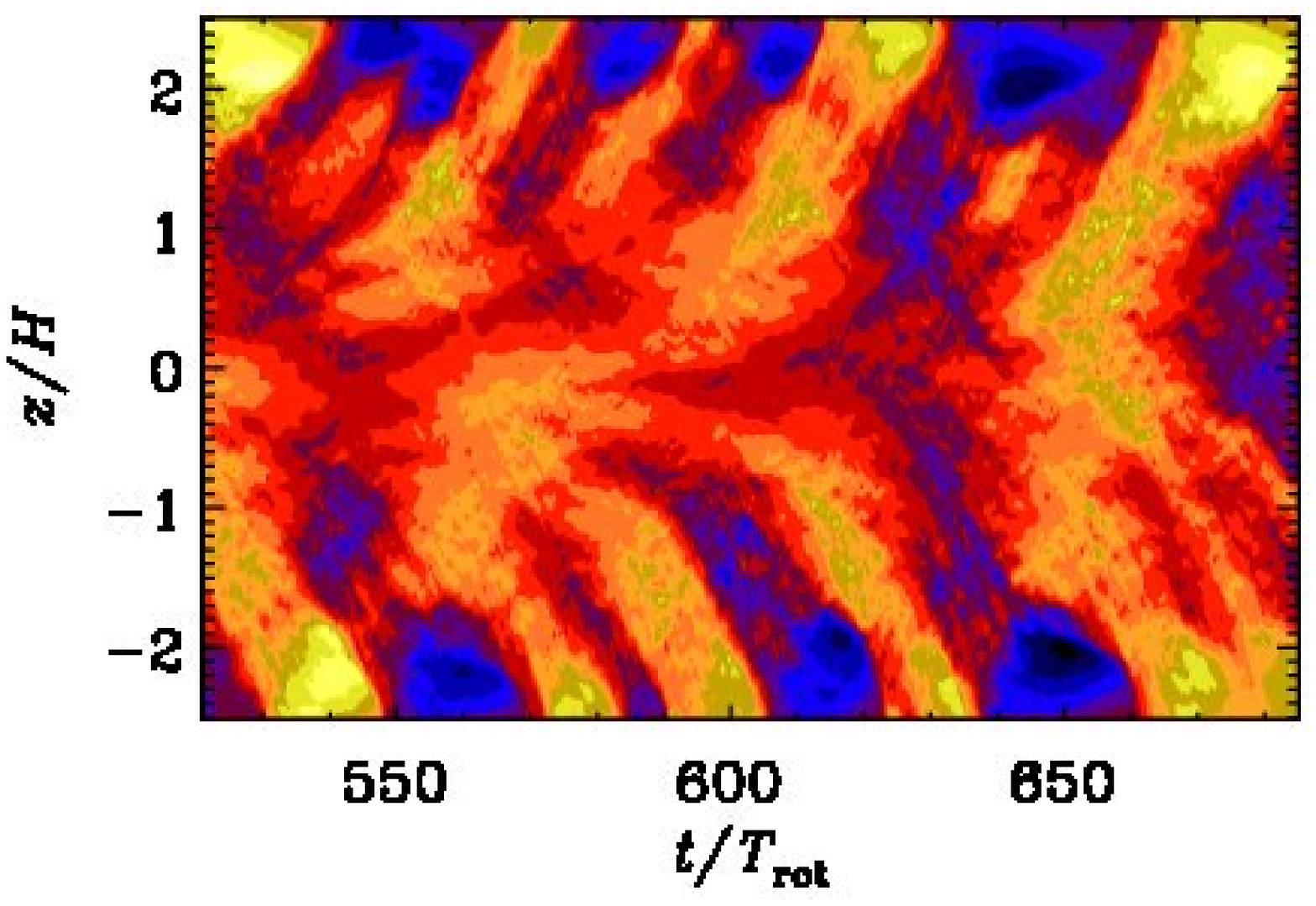}
\end{center}\caption[]{
Horizontally averaged magnetic fields $\mean{B}_x$ (upper
panel on the left) and $\mean{B}_y$ (lower panel on the left),
as functions of time and $z$ for a run with shear and no rotation
(K\"apyl\"a et al.\ 2008), as well as
$\mean{B}_y$ from a local model of an accretion disc (right hand panel),
where $z=0$ corresponds to the midplane.
On the left-hand panel the dotted white lines show top ($z=d$) and bottom
($z=0$) of the convection zone.}
\label{but256a}
\end{figure}

\section{Quenching and examples of large-scale dynamos}

Since the early 1990s there has been the notion that catastrophic
$\alpha$ quenching has been a major concern in dynamo theory
(Vainshtein \& Cattaneo 1992).
This led Parker (1993) to suggest that the solar dynamo might work in
a segregated fashion where shear acts only near the lower overshoot
layer where the turbulent magnetic diffusivity is low.
In the bulk of the convection zone, on the other hand,
the magnetic field is supposed to be weak, and so there was hope that
the $\alpha$ effect would not be catastrophically quenched.
However, only later the origin of catastrophic quenching was understood
to be due to magnetic helicity conservation.
As a consequence, catastrophic quenching could not be alleviated by
spatial rearrangements, but only by magnetic helicity fluxes out of
the domain.
Indeed, there are now a number of simulations that show successful
generation of large-scale magnetic fields; see \Fig{but256a}.
All these simulations have in common that there are contours of
constant shear crossing the outer surface.
This is believed to be important because the magnetic helicity flux
is expected to be directed along the contours of constant shear
(Brandenburg \& Subramanian 2005a).

\section{$\alpha$ effect and production of small-scale magnetic helicity}

\begin{figure}[t!]\begin{center}
\includegraphics[width=.32\textwidth]{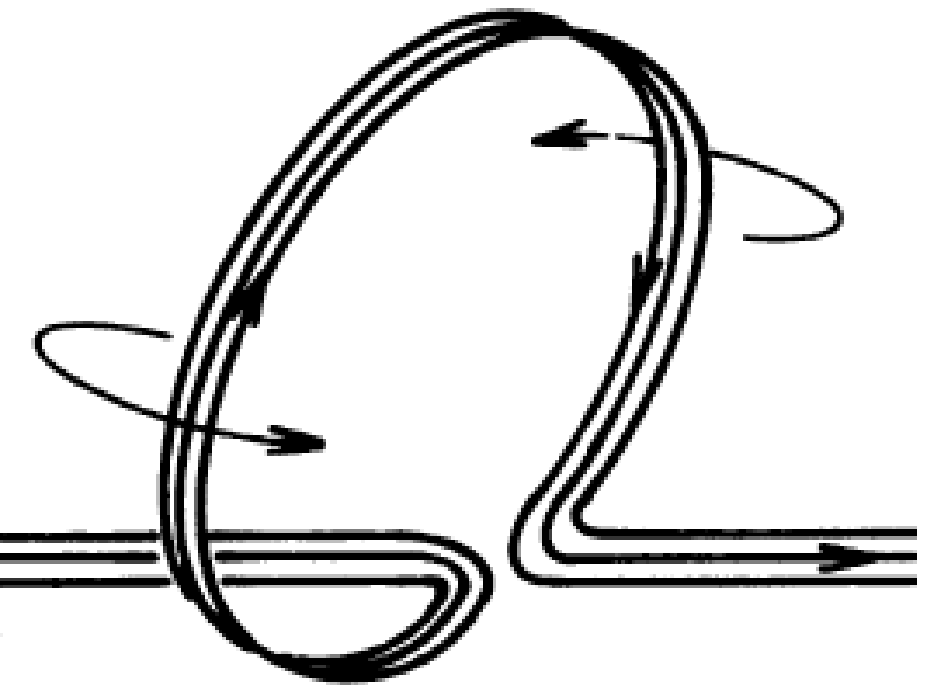}
\includegraphics[width=.32\textwidth]{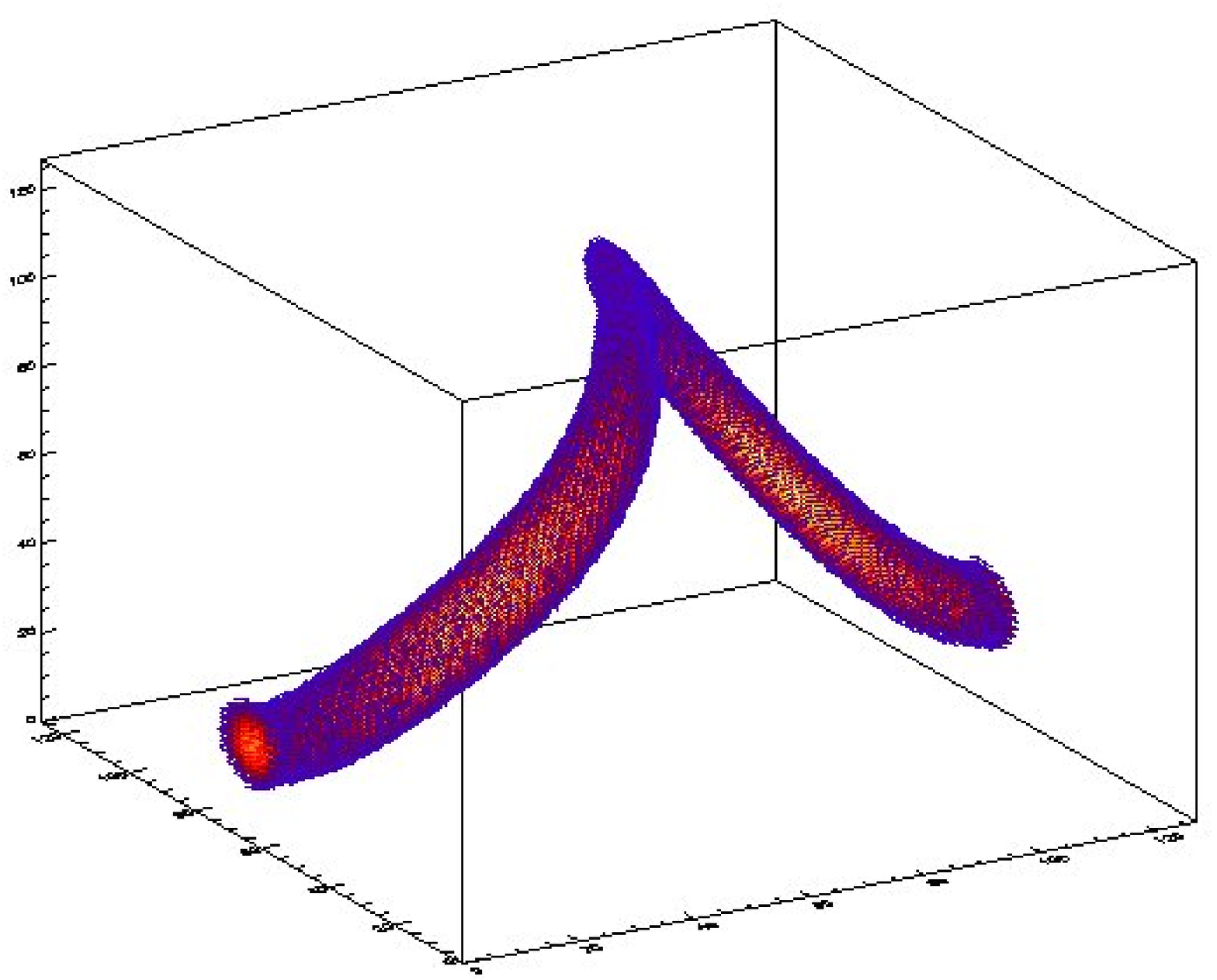}
\includegraphics[width=.32\textwidth]{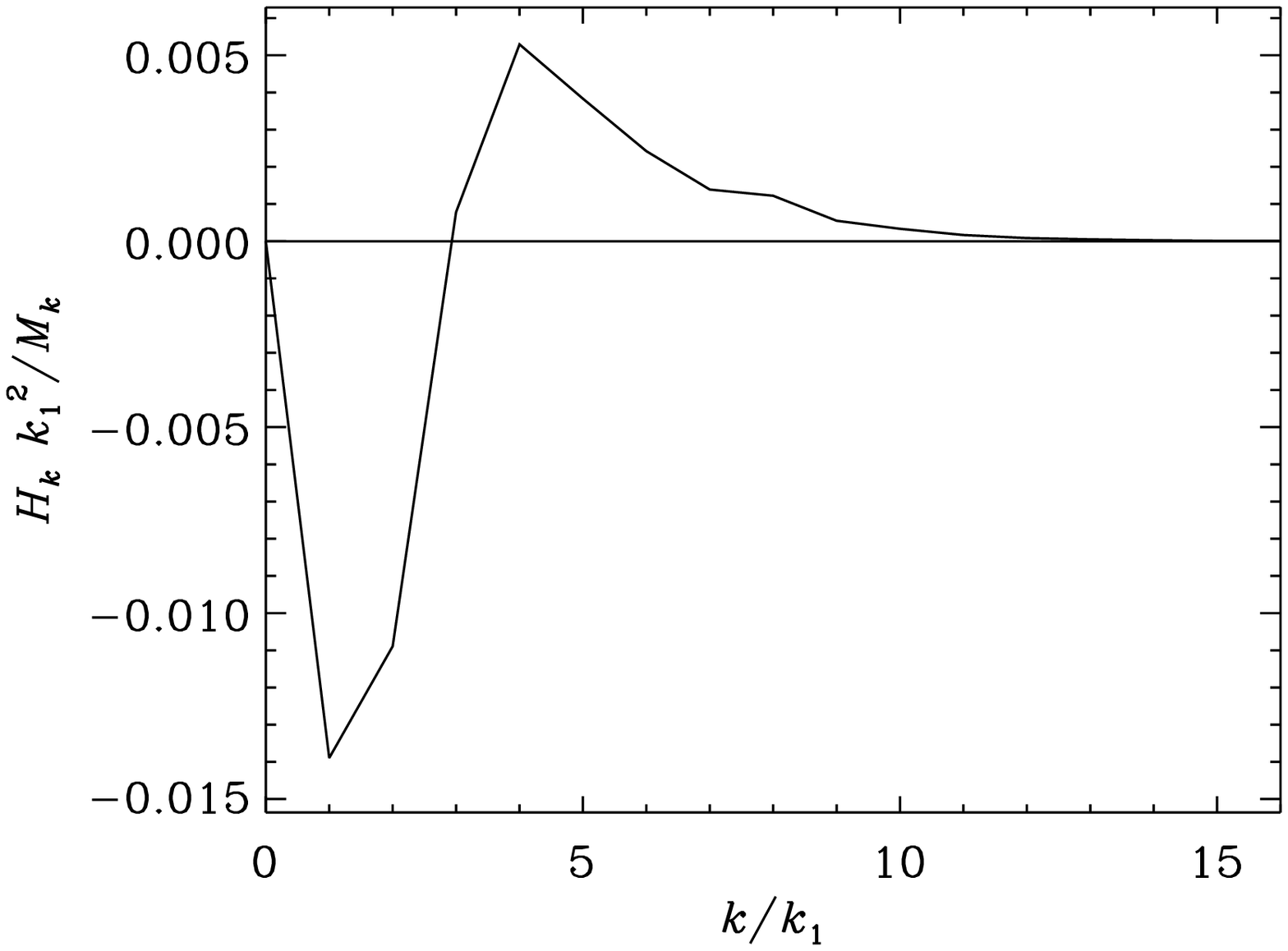}
\end{center}\caption[]{
Original sketch of a Parker loop (Parker 1957) on the left,
compared with a loop constructed from the Cauchy solution of
Yousef \& Brandenburg (2003) and its corresponding 
magnetic helicity spectrum $H_k$ on the right.
}\label{loop}
\end{figure}

The $\alpha$ effect is central to many mean-field approaches of the
solar dynamo.
Its presence is usually associated with the systematic action of the
Coriolis force on convecting fluid elements.
As such an element rises vertically, it expands and counter rotates due
to the Coriolis force, thereby attaining a downward oriented vorticity
in the northern hemisphere, and an upward oriented vorticity in the
southern hemisphere.
As the magnetic field is dragged along with the gas, a systematic
poloidal field component is being produced from a toroidal field.
An early sketch of this process was given by Parker (1957), where a rising
loop attains a systematic tilt; see the left hand panel of \Fig{loop}.

In recent years the nonlinear saturation of this process has been
understood to be due to the internal twist that is being induced as
the loop begins to tilt.
This is already quite evident from the early sketch by Parker (1957)
suggesting the presence of a spatially extended structure of the loop
with oppositely oriented twisting motions on both ends of the loop;
see the round arrows on the tube in the left hand panel of \Fig{loop}.

In the middle panel of \Fig{loop} we show a visualization of an
analytically generated representation of a Parker loop using the
Cauchy solution for ideal magnetohydrodynamics together with the
corresponding numerically constructed magnetic helicity spectrum
(Yousef \& Brandenburg 2003); see the right hand panel of \Fig{loop}.
This spectrum shows quite clearly a negative magnetic helicity at
low wavenumbers $k/k_1\approx1$ (where $k_1=2\pi/L$ is the smallest
wavenumber in a box of scale $L$) and positive magnetic helicity at
larger wavenumbers around $k/k_1\approx4$.
This confirms an early finding of Seehafer (1996) that the $\alpha$ effect
corresponds to a segregation of magnetic helicity in scale and that magnetic
helicity of smaller scale and opposite sign is inevitably being produced
by the $\alpha$ effect.
It is this what quenches the $\alpha$ effect in a potentially
catastrophic manner.
Alleviating this type of quenching requires that we have to get rid of
this smaller-scale magnetic helicity through helicity fluxes.
This is something that the Sun seems to accomplish through coronal
mass ejections (D\'emoulin et al.\ 2002, Blackman \& Brandenburg 2003).
In the following we describe in more detail the magnetic helicity
flux and its relation to the amount of shear.

\section{Magnetic helicity flux}

There are a number of analytic calculations of magnetic helicity fluxes
relevant to astrophysical dynamos (Kleeorin et al.\ 2000, 2002, 2003a,b;
Vishniac \& Cho 2001; Subramanian \& Brandenburg 2004, 2006;
Brandenburg \& Subramanian 2005a,b).
Here we present preliminary numerical calculations in a simple system.
We calculate the small-scale current helicity flux, $\meanFFFF_C^{\rm SS}$,
in homogeneous turbulence in the presence of shear, $S$, and a uniform
magnetic field $\BB_0$.
Since the system is completely uniform, it makes sense to consider full
volume averages, denoted here by an overbar.
The wavenumber of the averaged quantities is zero, so we have infinite
scale separation and can then force the turbulence at the scale of
the system, i.e.\ at wavenumbers between 1 and 2 times the smallest
finite wavenumber, $k_1$.
Our average forcing wavenumber is therefore $\kf/k_1=1.5$.
We consider the dependence on the parameters $B_0/\Beq$, $\Rm$ and $\Sh$, where
\EQ
\Beq^2=\mu_0\bra{\rho\uu^2},\quad
\Rm=\urms/\eta\kf,\quad
\Sh=S/\eta\kf^2.
\EN
The small-scale current helicity flux is calculated from the expression
(Brandenburg \& Subramanian 2005b)
\EQ
\meanFFFF_C^{\rm SS}=2\overline{\ee\times\jj}+\overline{(\nab\times\ee)\times\bb},
\label{FCSS}
\EN
where
\EQ
\ee=-\uu\times(\bb+\BB_0)+\eta\mu_0\jj
\EN
is the small-scale electric field and $\jj=\nab\times\bb/\mu_0$ is the
small-scale current density with $\mu_0$ being the vacuum permeability.
In order to avoid taking more than two derivatives we integrate
the second expression in \Eq{FCSS} by parts, i.e.\
\EQ
\overline{(\nab\times\ee)\times\bb}=\overline{(\nab\bb)^T\ee},
\label{FCSS2}
\EN
where we have used the fact that $\nab\cdot\bb=0$.
The $i$ component of this term can also be written as
$\overline{b_{j,i}e_j}$, where a comma denotes partial differentiation.

\begin{figure}[t!]\begin{center}
\includegraphics[width=.32\columnwidth]{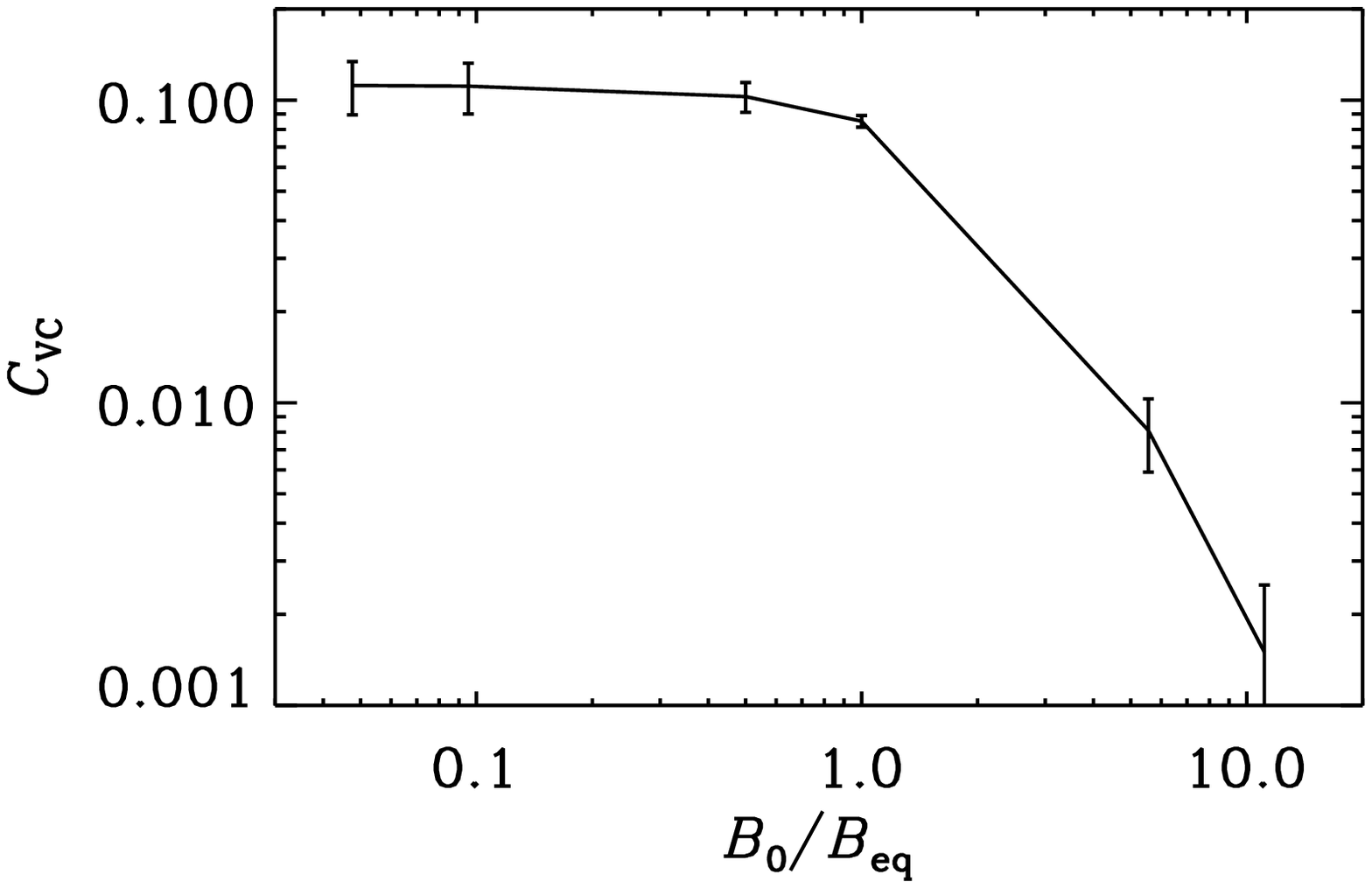}
\includegraphics[width=.32\columnwidth]{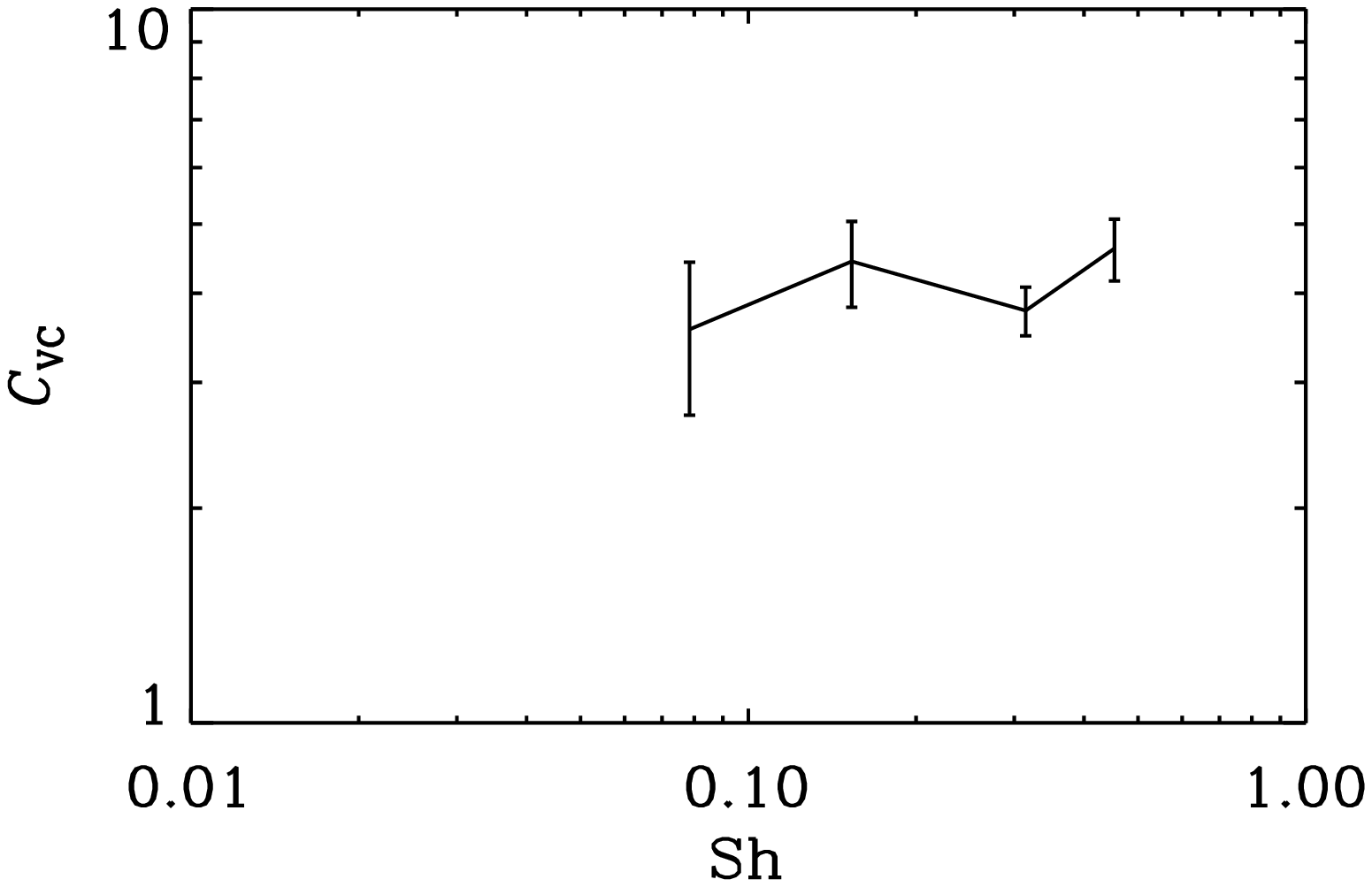}
\includegraphics[width=.32\columnwidth]{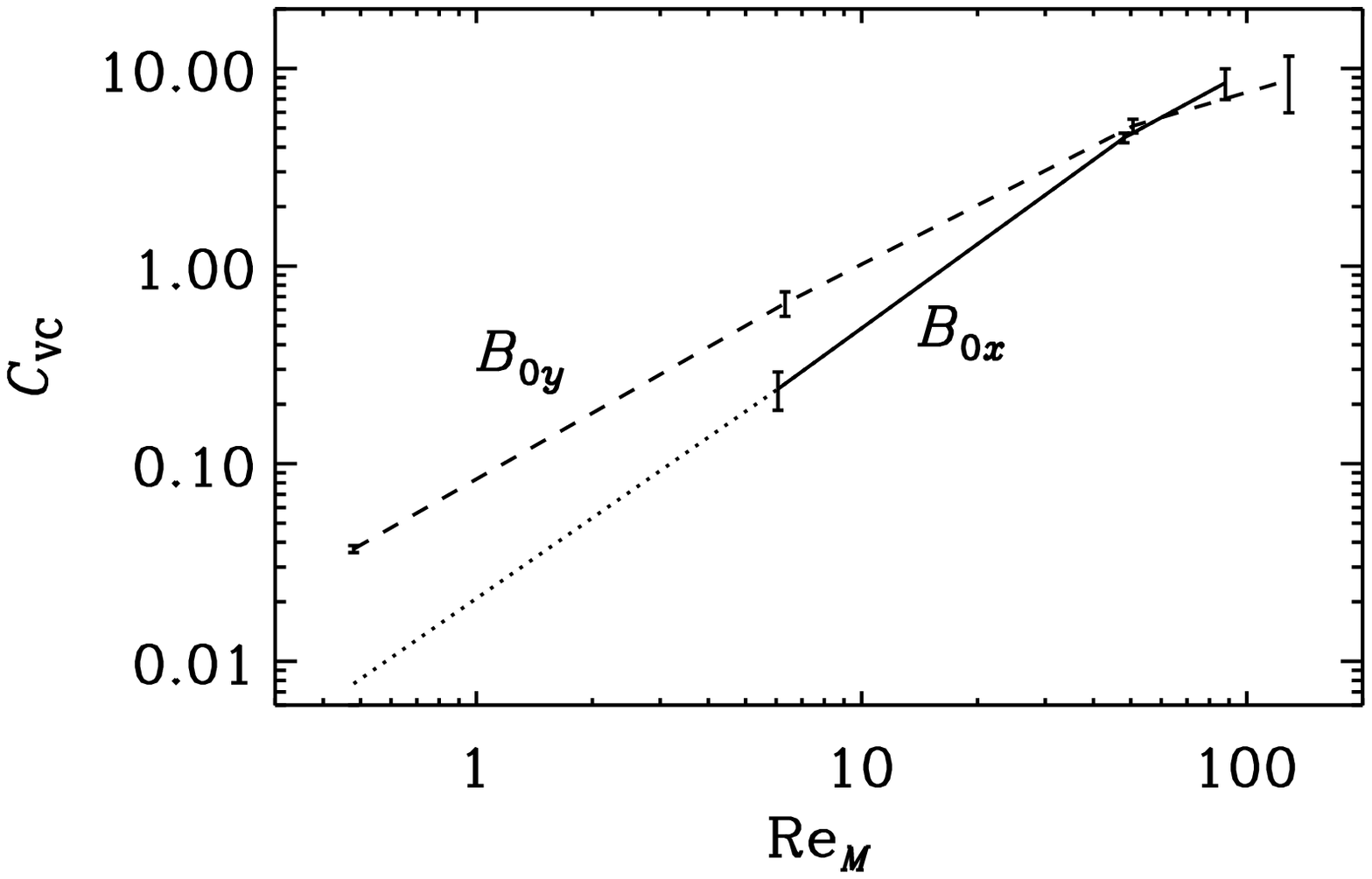}
\end{center}\caption[]{
$B$ dependence of $-C_{\rm VC}$ for $\BB_0=B_{0y}\yyy$ with $\Rm=1.4$,
$\Sh\approx-2$ (left),
$S$ dependence of $C_{\rm VC}$ for $\Rm=30$, $B_0/\Beq\approx0.2$ (middle), and
$\Rm$ dependence of $C_{\rm VC}$ for $\Sh\approx0.5$, $B_0/\Beq\approx0.2$.
}\label{pbdep}\end{figure}

Based on calculations using the minimal $\tau$ approximation
the small-scale current helicity flux is expected to be given by
(Brandenburg \& Subramanian 2005a)
\EQ
\meanFFFF_C^{\rm SS}=C_{\rm VC}(\meanSSSS\meanBB)\times\meanBB,
\EN
where $C_{\rm VC}$ is a non-dimensional coefficient that is of the
order of $\St^2$, where $\St=\tau\urms\kf$ is the Strouhal number.
Throughout this work we use a uniform shear flow, i.e.\
$\meanUU=(0,Sx,0)$, so the $z$ component of $\meanFFFF_C^{\rm SS}$ is
\EQ
\meanFFF_{Cz}^{\rm SS}=\half C_{\rm VC}S\left(\meanB_y^2-\meanB_x^2\right).
\EN
In the following we quote values of $C_{\rm VC}$ that are computed
by imposing a uniform field either in the $x$ or in the $y$ direction.
The numerical resolution is only $32^3$.
In the left hand panel of \Fig{pbdep} we show the dependence on $B_0$
for a small value of $\Rm$ ($\Rm=1.4$) and fixed shear parameter
$\Sh\approx-2$ (for $S<0$).
So far there is no indication that the flux depends on $\Sh$; see
the middle panel of \Fig{pbdep}.
However, the flux shows a strong increase with $\Rm$; see the right
hand panel of \Fig{pbdep}.
It should be noted that, since we keep the forcing unchanged while
changing the viscosity, the resulting rms velocity also changes, and
so the resulting values of $\Sh$ and $B_0/\Beq$ also change somewhat.

\section{Discussion}\label{sec:concl}

In this paper we have presented selected aspects of the solar dynamo
problem where there has been recent progress.
The idea that the solar dynamo may operate in the bulk of the convection
zone is motivated in part by the fact that it is easier to dispose of
small-scale magnetic helicity from upper layers than from deeper down.
Also the sign of the radial differential rotation is negative and would
produce equatorward migration of dynamo waves in the presence of a
positive $\alpha$ effect in the northern hemisphere of the Sun.
A number of other arguments for a distributed solar dynamo, where the
field that makes sunspots does not solely come from the lower overshoot
layer, have been discussed elsewhere (Brandenburg 2005).
However, what is not yet well addressed is the production of sunspots
within active regions, and perhaps even the active regions themselves.
It is plausible, and it has indeed been argued, that sunspots can be
the result of negative turbulent magnetic pressure effects
(Kleeorin \& Rogachevskii 1994) or a turbulent magnetic collapse
phenomenon (Kitchatinov \& Mazur 2000).
Both processes rely on turbulent transport processes that could be
verified numerically and whose effects could also be demonstrated
directly in suitably arranged simulations.

\begin{acknowledgments}
The computations have been carried out at the National Supercomputer
Centre in Link\"oping and at the Center for Parallel Computers at the
Royal Institute of Technology in Sweden.
This work was supported in part by the Swedish Research Council.
\end{acknowledgments}

\begin{discussion}
\discuss{Strassmeier}{Would a small (ideal) isolated spot trace the
large-scale or the small-scale field?}
\discuss{Brandenburg}{Large-scale field means really mean field, and
if we define this through azimuthal averaging, not much of the spot
field would survive azimuthal averaging. Therefore, most of the spot field
would contribute to the small-scale field and only a small fraction
would contribute to the large-scale field.}
\end{discussion}
\end{document}